\documentclass[aps,prc,reprint,groupedaddress]{revtex4-1}
\usepackage{amsmath}
\usepackage[dvips]{graphicx}

\begin{document}

%\preprint{}

\title{One dimensional hydrodynamical model including phase transition}


\author{Naomichi \textsc{Suzuki}}
\email[]{suzuki@matsu.ac.jp}
%\thanks{}
\affiliation{Department of Comprehensive Management, Matsumoto University, 
 Matsumoto 390-1295, Japan
}


\date{\today}

\begin{abstract}
 Analytical solution of one dimensional hydrodynamical model is derived, where
phase transition from the QGP state to the hadronic state is effectively 
taken into account.
The single particle rapidity distribution of charged $\pi$ mesons observed 
in relativistic heavy ion collisions is analyzed by the model. 
Space-time development of the fluid is also investigated. 
\end{abstract}

%\keywords{}

\maketitle

\section{Introduction}

%\subsection{}
%\subsubsection{}

It is widely recognized that at the initial stage of relativistic heavy ion collisions, the quark-gluon plasma (QGP) is formed, 
and it breaks up to thousand of hadrons in the final states. 
It is also considered that hydrodynamical approach is one of the effective tools~\cite{land53, khal54,bele56} 
to analyze such multiple particle production processes.

In most of numerical calculations of three dimensional hydrodynamical models, 
the phase transition from the QGP state to the hadronic state is taken into account~\cite{nona00, mori02, kolb04}.  
 For reviews of the recent works, see \cite{hira08}.

Recently, exact solutions of one dimensional hydrodynamical model are investigated~\cite{bial07,nagy08,beuf08,mizo09}. 
However, the phase transition from the QGP state is not introduced to the analytical formulations. 

The hydrodynamical model is consisted of the energy momentum conservation, 
the baryon number conservation (or entropy conservation) and the equation of state.   
The phase transition from the QGP state to the hadronic state can be expressed by the change of the velocity of sound 
from that of the perfect fluid, $c_0=1/\sqrt{3}$, to some constant value, $c_s$, at critical temperature. 
We assume that shear viscosity, bulk viscosity and heat conductivity are negligibly small in the hadronic state. 
We also assume baryon number can be neglected.
 
One dimensional hydrodynamical model of the perfect fluid can be solved with an appropriate initial condition. 
If the change of the velocity of sound is taken into the model, 
we can formulate the hydrodynamical model including the phase transition from the QGP state to the hadronic state.

In section \ref{ODHM}, one dimensional hydrodynamical model following to Landau's approach is reviewed, 
and a solution for potential $\chi$ is derived under an idealized initial condition. 
In section \ref{HDPT}, one dimensional hydrodynamical model including the phase transition from the QGP state to the hadronic state is formulated. 
In section \ref{SPRD}, the single particle distribution under the Cooper-Frye approach is shown. 
The single particle rapidity distribution observed in relativistic nucleus-nucleus (AA) collisions is analyzed in section \ref{DA}. 
The space-time evolution of fluid element is also investigated there. The final section is devoted to concluding remarks.

\section{One dimensional hydrodynamical model}
\label{ODHM}

\subsection{Equations for hydrodynamical model}
The hydrodynamical model proposed by Landau~\cite{land53, khal54,bele56} is composed from the 
energy-momentum conservation of the perfect fluid and the equation of state.  

The energy-momentum conservation is given by, 
 \begin{eqnarray}
   \frac{\partial}{\partial x^\nu} {T^{\mu\nu}} &=& 0,  \quad \mu, \nu=0, 1,   \label{eq.odm1}  
 \end{eqnarray}
where $T_{\mu\nu}$ denotes the energy momentum tenser of the perfect fluid, 
 \begin{eqnarray}
   T_{\mu\nu} = (\varepsilon + p) u_{\mu}u_{\nu} - p g_{\mu\nu}.   \label{eq.odm1b}  
 \end{eqnarray}
In Equation (\ref{eq.odm1b}), $\varepsilon$ denotes the energy density and $p$ the pressure of a fluid element. 
The velocity of the fluid element is denoted by $u^\mu$, which satisfies, $u^\mu u_\mu =1$, 
and $g^{\mu\nu} = {\rm diag}(1, -1)$. 

The equation of state is expressed as a function of the energy density $\varepsilon$ to the pressure $p$. 
It would be different whether the fluid is in the QGP state or in the hadronic state. 
At present we assume that the relation, ${dp}/{d\varepsilon} =  {c_0}^2$, holds,  
where  $c_0$ is a positive constant.

In addition, thermodynamical relations, 
 \begin{eqnarray}
     \varepsilon + p = Ts,  \quad d\varepsilon = T ds,        \label{eq.odm3}
 \end{eqnarray}
are used, where $T$ denotes the temperature and $s$ is the entropy density of the fluid element.  
Equation (\ref{eq.odm3}) holds whether fluid is in the QGP state or in the hadronic state.

Projection of Eq.(\ref{eq.odm1}) to the direction of $u_\mu$ gives the entropy conservation, 
 \begin{eqnarray}
     \frac{\partial (s u^\nu)}{\partial x^\nu} = 0.          \label{eq.odm4}
 \end{eqnarray}
After Eq.(\ref{eq.odm1}) is projected to the direction perpendicular to $u_\mu$, we obtain,
 \begin{eqnarray} 
    -\frac{\partial T}{\partial x^\mu} +  \frac{\partial (Tu_\nu)}{\partial x^\mu} u^\nu = 0, \quad \mu=0, 1. 
           \label{eq.odm7}       
 \end{eqnarray}
Rapidity $\eta$ of the fluid element is defined by, 
 \begin{eqnarray} 
    (u^0, u^1) = ( \cosh \eta , \sinh \eta ).    \label{eq.odm8}
 \end{eqnarray}
Then, both equations in Eq.(\ref{eq.odm7}) reduce to
 \begin{eqnarray} 
       \frac{\partial}{\partial t}( T \sinh \eta )
         + \frac{\partial}{\partial x}( T \cosh \eta ) = 0.       \label{eq.odm9}              
 \end{eqnarray}
From Eq.(\ref{eq.odm9}), there is a function, $\phi$, which satisfies,
 \begin{eqnarray}
     \frac{\partial \phi}{ \partial t} = T\cosh  \eta , \quad
     \frac{\partial \phi}{ \partial x} = -T\sinh  \eta.          \label{eq.vp1}
 \end{eqnarray}
By the use of the Legendre transform,
 \begin{eqnarray*}
     d\chi = d(\phi - tT \cosh \eta + xT \sinh \eta ),  \nonumber             
 \end{eqnarray*}
we obtain the equations for the potential $\chi$~\cite{khal54,bele56} ; 
 \begin{eqnarray}
    \frac{\partial \chi}{\partial T}&=& -t\cosh \eta + x\sinh \eta ,   \nonumber \\
    \frac{1}{T}\frac{\partial \chi}{\partial \eta} &=& -t\sinh \eta + x\cosh \eta.  \label{eq.vp2}          
 \end{eqnarray}

Then, space-time variables of the fluid element, $t$ and $x$, are given respectively as，
 \begin{eqnarray}
    t &=& \frac{1}{T_0}{\rm e}^\omega\Big( \frac{\partial \chi}{\partial \omega}\cosh \eta 
         + \frac{\partial \chi}{\partial \eta }\sinh \eta \Big),   \nonumber \\
    x &=& \frac{1}{T_0}{\rm e}^\omega\Big( \frac{\partial \chi}{\partial \omega}\sinh \eta 
         + \frac{\partial \chi}{\partial \eta }\cosh \eta \Big),   \label{eq.vp3}        
 \end{eqnarray}
where $T_0$ is the initial temperature  of the fluid and $\omega=\ln(T_0/T)$.

The entropy conservation, Eq.(\ref{eq.odm4})，is rewritten as,
 \begin{eqnarray}
    \frac{\partial }{\partial t} (s\cosh  \eta ) + 
    \frac{\partial }{\partial x} (s\sinh  \eta ) = 0.       \label{eq.vp4}    
 \end{eqnarray}
After changing the variables from $x,t$ to $\eta, T$ in  Eq.(\ref{eq.vp4}), we obtain,
 \begin{eqnarray*}
     \frac{T}{s}\frac{ds}{dT}\Big( 
        \frac{\partial^2 \chi}{\partial \eta ^2} - T\frac{\partial \chi}{\partial T}\Big)
         - T^2 \frac{\partial^2 \chi}{\partial T^2}  = 0.   
 \end{eqnarray*}

\subsection{Solution for potential $\chi$}

Using the relation $(T/s)ds/dT={c_0}^2$ and the variable, $\omega=-\ln (T/T_0)$,
 we obtain the partial differential equation for the potential $\chi$;
 \begin{eqnarray}
   && \frac{\partial^2 \chi}{\partial\omega^2}
         - 2\beta_0 \frac{\partial \chi}{\partial\omega}-\frac{1}{{c_0}^2} \frac{\partial^2 \chi}{\partial \eta ^2}= 0,  \label{eq.vp5}
  \\
   &&        \beta_0=\frac{1-{c_0}^2}{2{c_0}^2}=1.  \nonumber  \end{eqnarray}
Equation (\ref{eq.vp5}) is called the equation of telegraphy.

The initial condition for Eq.(\ref{eq.vp5}) is taken with a constant $Q_0$ as,
 \begin{eqnarray}
      \chi|_{\omega=0} = g(\eta) = 0, \quad  
       \frac{\partial \chi}{\partial\omega}\Big|_{\omega=0} = G(\eta) = Q_0\delta(\eta).   \label{eq.vp6}  
 \end{eqnarray}
Hydrodynamical model is applicable mainly to the central region~\cite{nami57}. 
As can be seen from Eq.(\ref{eq.vp10}),  
the main contribution to the fragmentation comes from the function  $g(\eta)$. 
Therefore, we assume that $g(\eta)=0$. At the initial stage, fluid would be formed in the very small region of rapidity space, 
and we put $G(\eta)$ is proportional to the delta function in the rapidity space as an idealized case~\cite{suzu08}. 

Introducing the new variable $\chi_1$ by,
 \begin{eqnarray} 
      \chi(\eta, \omega) =  \chi_1(\eta, \omega){\rm e}^{\beta_0\omega},  \label{eq.vp7}
 \end{eqnarray}
we have the partial differential equation for $\chi_1$  as, 
 \begin{eqnarray} 
     \frac{\partial^2 \chi_1}{\partial\omega^2} - {\beta_0}^2\chi_1
          - \frac{1}{{c_0}^2}\frac{\partial^2 \chi_1}{\partial \eta^2}= 0,      \label{eq.vp8} 
 \end{eqnarray}
and the initial condition for Eq.(\ref{eq.vp8}) as,
 \begin{eqnarray}
      \chi_1|_{\omega=0} &=& g(\eta) = 0,  \nonumber  \\
      \frac{\partial \chi_1}{\partial\omega} \Big|_{\omega=0} &=& G(\eta)-\beta_0 g(\eta) = Q_0 \delta(\eta). 
        \label{eq.vp9}
 \end{eqnarray}
The solution for $\chi_1$ is given in general as~\cite{kosh74},
%
 %\begin{widetext}
  \begin{eqnarray}
       &&  \chi_1(\eta, \omega) = \frac{1}{2} \{ g(\eta+\omega/c_0) + g(\eta - \omega/c_0) \} \nonumber \\
               && \hspace{5mm}  + \frac{c_0}{2} \int_{-\omega/c_0}^{\omega/c_0}dz \, 
                                      \Bigl\{ G(z+\eta)-\beta_0 g(\eta) \Bigr\}   \nonumber \\
             && \hspace{25mm}    \times 
             I_0\Bigl(\beta_0\sqrt{\omega^2-{c_0}^2z^2}\, \Bigr), \nonumber \\ 
       && \hspace{5mm}   + \frac{\beta_0c_0\omega}{2} \int_{-\omega/c_0}^{\omega/c_0}dz\, g(z+\eta)    \nonumber \\
       && \hspace{25mm} \times  \frac{ I_1(\beta_0\sqrt{\omega^2-{c_0}^2z^2}\,)} {\sqrt{\omega^2-{c_0}^2z^2} }.     \label{eq.vp10} 
  \end{eqnarray}
 %\end{widetext}
%
It reduces to                                      
 \begin{eqnarray}                    
    \chi_1(\eta,\omega) = \frac{Q_0 c_0}{2} 
         I_0\bigl(\beta_0\sqrt{\omega^2-{c_0}^2 \eta ^2}\,\Bigr),   \label{eq.vp11}  
 \end{eqnarray}
for $|\eta|<\omega/c_0$.

Then the solution for Eq.(\ref{eq.vp5}) under the initial condition (\ref{eq.vp6}) is given by,
 \begin{eqnarray}
        \chi(\eta, \omega) =  
          \frac{Q_0 c_0}{2}{\rm e}^{\beta_0\omega}I_0\Bigl(\beta_0\sqrt{\omega^2-{c_0}^2 \eta^2}\,\Bigr).
                   \label{eq.vp12}
 \end{eqnarray}
\section{Hydrodynamical model including phase transition}
\label{HDPT}

\subsection{Introduction of phase transition}

In our formulation, the phase transition from the QGP state to the hadronic state is expressed 
by the change of the velocity of sound. 
The evolution of fluid is described by the parameter $\omega=\ln(T_0/T)$, where $T_0$ is the initial temperature. 
The fluid formed just after the collision of nuclei, is assumed to be in the QGP state. 
The velocity of sound in it is denoted by $c_0$ ($c_0=1/\sqrt{3}$).
It expands from $\omega = 0$ to $\omega = \omega_c-0$, where $\omega_c =\ln(T_0/T_c)>0$ with the critical temperature $T_c$.  
The phase transition from the QGP to the hadronic state occurs at $\omega=\omega_c$. 
From $\omega=\omega_c+0$ to $\omega = \omega_f$, where $\omega_f=\ln(T_0/T_f)>\omega_c$ with the freeze-out temperature $T_f$,  
the fluid is in the hadronic state. The velocity of sound in it is assumed to be constant and denoted by $c_s$ ($0<c_s \le c_0$).

\begin{figure}[!htb]
   \begin{center}
      \includegraphics[width=6.cm,clip]{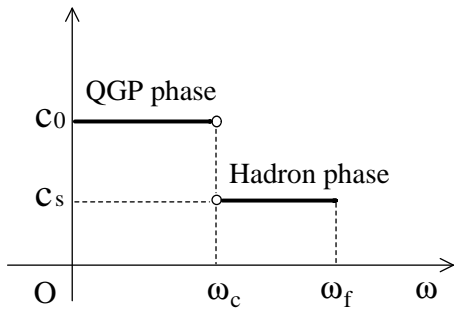}
   \end{center}
    \caption{\label{fig.vs}the velocity of sound as a function of $\omega$ from  $\omega=0$ to  $\omega=\omega_f$}
 \end{figure}

 The equation of state is assumed to be,
 \begin{eqnarray} 
     p = \begin{cases}
          {c_0}^2 \varepsilon - (1+{c_0}^2)B, & \text{for}\,\, 0 < \omega < \omega_c,   \\
          {c_s}^2 \varepsilon,  &  {\rm for}\,\, \omega_c < \omega < \omega_f,        \label{eq.qgp0}
        \end{cases}
 \end{eqnarray}
where $B$ is the bag constant of hadrons.
From Eq.(\ref{eq.qgp0}), the velocity of sound is given by
 \begin{eqnarray}
    \sqrt{\frac{dp}{d\varepsilon}} = 
              \begin{cases}
                 {c_0}, &  {\rm for}\,\, 0 < \omega < \omega_c,   \\
                 {c_s}, &  {\rm for}\,\, \omega_c < \omega < \omega_f.  
              \end{cases}            
                               \label{eq.qgp1a}   
 \end{eqnarray}
As is shown in Figure \ref{fig.vs}, the velocity of sound is discontinuous at $\omega=\omega_c$.

\subsection{Potential $\chi(\eta, \omega)$ in the QGP fluid}

We assume that the fluid in the QGP phase, where the velocity of sound is $c_0 = 1/\sqrt{3}$,  
continues from $\omega=0$ to $\omega=\omega_c-0$. 
Then, the partial differential equation for the potential $\chi(\eta, \omega)$ is given by Eq.(\ref{eq.vp5});
 \begin{eqnarray}
    \frac{\partial^2 \chi}{\partial\omega^2}
         - 2\beta_0 \frac{\partial \chi}{\partial\omega}
         -\frac{1}{{c_0}^2} \frac{\partial^2 \chi}{\partial \eta^2}= 0, 
           \label{eq.qgp1}
 \end{eqnarray}
where  $\beta_0 = {(1-{c_0}^2)}/{2{c_0}^2}=1$. 
The initial condition for Eq.(\ref{eq.qgp1}) is taken as Eq.(\ref{eq.vp6}),
 \begin{eqnarray}
      \chi|_{\omega=0} = g(\eta) = 0, \quad  
       \frac{\partial \chi}{\partial\omega}\Big|_{\omega=0} = G(\eta) = Q_0\delta(\eta).   \label{eq.qgp2}  
 \end{eqnarray}

The solution for Eq.(\ref{eq.qgp1}) under the initial condition (\ref{eq.qgp2}) is given by,
 \begin{eqnarray}
        \chi(\eta, \omega) = \frac{Q_0 c_0}{2}{\rm e}^{\beta_0\omega}I_0\Bigl(\beta_0\sqrt{\omega^2-{c_0}^2 \eta^2}\,\Bigr).
                   \label{eq.qgp8}
 \end{eqnarray}
\subsection{Potential $\chi(\eta, \omega)$ in the hadronic fluid}

In the hadronic phase, the fluid expands from $\omega=\omega_c+0$ to  $\omega=\omega_f$. 
During the expansion of the hadronic fluid, 
with the velocity of sound is $c_s$. 

For $\omega_c<\omega<\omega_f$, the partial differential equation for the potential $\chi$ is given by,
 \begin{eqnarray}
    \frac{\partial^2 \chi}{\partial\omega^2}- 2\beta \frac{\partial \chi}{\partial\omega}
          -\frac{1}{{c_s}^2} \frac{\partial^2 \chi}{\partial \eta^2}= 0, 
      \quad  \beta=\frac{1-{c_s}^2}{2{c_s}^2}\ge 1.   \label{eq.hdf1}
 \end{eqnarray}
For $|\eta|<(\omega-\omega_c)/c_s +\omega_c/c_0$，the solution for  Eq.(\ref{eq.hdf1}) is given by
 \begin{eqnarray}
     \chi(\eta,\omega) &=& A(\omega)I_0(\lambda), \quad  \nonumber  \\
         A(\omega) &=& \frac{Q_0c_0}{2}{\rm e}^{\beta(\omega-\omega_c)+\beta_0\omega_c},\label{eq.hdf2} 
 \end{eqnarray}
where
 \begin{eqnarray}
     \lambda &=& \beta c_s \sqrt{\eta_{max}^2-\eta^2}, \quad   \nonumber  \\
     \eta_{max} &=& (\omega-\omega_c)/c_s +\omega_c/c_0.   \label{eq.hdf3}      
 \end{eqnarray}

If $c_s=c_0$ ($\beta=\beta_0$)，the potential $\chi(\eta, \omega)$ in Eq.(\ref{eq.hdf2}) coincides with Eq.(\ref{eq.qgp8}), 
which is the solution for Eq.(\ref{eq.qgp1}) under the initial condition (\ref{eq.qgp2}) ．

\subsection{Effect of phase transition on potential $\chi(\eta, \omega)$}

In our formulation, the velocity of sound is changed from $c_0$ to $c_s$ according to the phase transition 
from the QGP state to the hadronic state. 
This effect would appear in the potential $\chi(\eta, \omega)$ in the neighborhood of $\omega=\omega_c$.

At $\omega=\omega_c-0$, where the fluid is in the QGP state, the potential is given from Eq.(\ref{eq.qgp8}) as, 
 \begin{eqnarray}
        \chi(\eta, \omega_c-0) 
             = \frac{Q_0 c_0}{2}{\rm e}^{\beta_0\omega_c}I_0\Bigl(\beta_0\sqrt{{\omega_c}^2-{c_0}^2 \eta^2}\,\Bigr).
                   \label{eq.eft1}
 \end{eqnarray}
At $\omega=\omega_c + 0$, where the fluid is in the hadronic state, 
it is given from Eqs.(\ref{eq.hdf2}) and (\ref{eq.hdf3}) as,
 \begin{eqnarray}
     \chi(\eta,\omega_c+0) = \frac{Q_0c_0}{2}{\rm e}^{\beta_0\omega_c} 
        I_0\Bigr(\frac{\beta c_s}{c_0} \sqrt{{\omega_c}^2-{c_0}^2\eta^2} \Bigr).             \label{eq.eft2}      
 \end{eqnarray}
From Eqs.(\ref{eq.eft1}) and (\ref{eq.eft2}), we obtain,
 \begin{eqnarray}
    \frac{ \chi(\eta,\omega_c+0) }{ \chi(\eta, \omega_c-0) }
     = \frac{ I_0\Bigr(\frac{\beta c_s}{c_0} \sqrt{{\omega_c}^2-{c_0}^2\eta^2} \Bigr) }
        { I_0\Bigr(\beta_0 \sqrt{{\omega_c}^2-{c_0}^2\eta^2} \Bigr) } \neq 1.   \label{eq.eft3}      
 \end{eqnarray}

The potential $\chi(\eta, \omega)$ becomes discontinuous at $\omega=\omega_c$ due to the phase transition. 
This fact will affect the space-time behavior of fluid element. 

\section{Single particle rapidity distribution} 
\label{SPRD}

In order to analyze the rapidity distribution of observed particles, the Cooper-Frye approach~\cite{coop74} is used. 
Then, the single particle rapidity distribution $dn/dy$ of particles with mass $m$ and transverse momentum $p_T$ 
in nucleus-nucleus collisions is given by，
 \begin{eqnarray}
     \frac{dn}{dy} &=&  \frac{\pi {R_A}^2}{(2\pi)^3}
        \int_{\sigma}  \Bigl(  \cosh y \frac{dx}{d\eta} - \sinh y \frac{dt}{d\eta} \Bigr)\Big|_{\omega=\omega_f} \nonumber \\
      &&\times  \frac{m_T}{\exp[m_T \cosh(y-\eta)/T_f] - 1} d\eta d^2p_T,       \label{eq.opd1}
 \end{eqnarray}
where $R_A$ denotes the radius of colliding nuclei and $m_T = \sqrt{{p_T}^2 + m^2}$ denotes 
the transverse mass of observed particles.
Space-time variables, $t$ and $x$, in Eq.(\ref{eq.opd1}) is given by Eq.(\ref{eq.vp3}).
Under the assumption that the Bose-Einstein distribution can be approximated by the Maxwell-Boltzmann distribution, 
Eq.(\ref{eq.opd1}) is written as,
 \begin{widetext}
  \begin{eqnarray}
    &&  \frac{dn}{dy}= \frac{ {R_A}^2{T_f}^2}{4\pi} \int_{-\eta_{max}}^{\eta_{max}} d\eta
        \Bigl[ -\frac{\partial}{\partial \eta} \bigl( \chi + \frac{\partial \chi}{\partial\omega} 
                                          \bigr)\Big|_{\omega=\omega_f}\tanh (y-\eta) 
            +{c_s}^2 \frac{\partial}{\partial \omega} \bigl( \chi + \frac{\partial \chi}{\partial\omega} 
                                          \bigr)\Big|_{\omega=\omega_f} \Bigr]  \nonumber \\
    && \qquad \times\Bigl( \frac{m^2}{{T_f}^2} +\frac{2m}{T_f\cosh(y-\eta)}+\frac{2}{\cosh^2(y-\eta)}\Bigr)
       {\rm e}^{-m\cosh(y-\eta)/T_f}.   \label{eq.opd2}
  \end{eqnarray}
 \end{widetext}
If the term multiplied by $\tanh(y-\eta)$ is neglected in the square bracket on the right hand side of Eq.(\ref{eq.opd2}), 
the single particle rapidity distribution coincides with the Milekhin's approach~\cite{mile59,gore84} except for the normalization factor.

By the use of Eqs.(\ref{eq.hdf2}) and (\ref{eq.hdf3}), the single particle distribution is given by,
 \begin{widetext}
  \begin{eqnarray}
    \frac{dn}{dy}&=& \frac{{R_A}^2{T_f}^2}{4\pi}(\beta c_s)^2A(\omega_f)
                      \int_{-\eta_{max}}^{\eta_{max}} d\eta\, 
       \Bigl[  \frac{\beta \eta}{\lambda}
         \Bigl\{  \frac{\beta c_s\eta_{max}}{\lambda}{I_0}(\lambda) 
             +\Bigl(\frac{\beta + 1}{\beta} - 2\frac{\beta c_s\eta_{max}}{\lambda^2}\Bigr)I_1(\lambda) \Bigr\}                             
                                        \tanh (y-\eta)  \nonumber \\
   && \qquad  + \Bigl\{
               \Bigl(\frac{1 + \beta}{\beta} + \frac{(\beta c_s\eta_{max})^2}{\lambda^2}\Bigr) I_0(\lambda) 
             + \frac{\beta}{\lambda} 
            \Bigl( \frac{\eta_{max}}{c_s} + 1 - 2\frac{(\beta c_s\eta_{max})^2}{\lambda^2} \Bigr) I_1(\lambda) \Bigr\}\Bigr]  \nonumber \\
    && \qquad\quad \times\Bigl( \frac{m^2}{{T_f}^2} +\frac{2m}{T_f\cosh(y-\eta)}+\frac{2}{\cosh^2(y-\eta)}\Bigr)
       {\rm e}^{-m\cosh(y-\eta)/T_f}. 
              \label{eq.opd3} 
  \end{eqnarray}
 \end{widetext}
Space-time variables of the fluid element are written from Eqs.(\ref{eq.vp3}) and (\ref{eq.hdf2}) as,
 \begin{eqnarray}
    t &=& \frac{1}{T_0}A(\omega){\rm e}^\omega\Big[ \beta \Bigl\{ I_0(\lambda)
          + \beta c_s\frac{\eta_{max}}{\lambda}I_1(\lambda) \Bigr\}\cosh \eta   \nonumber \\
         &&\hspace{1.8cm}   -(\beta c_s)^2  \frac{\eta}{\lambda}I_1(\lambda)\sinh \eta \Big],  \label{eq.opd4}  \\
    x &=& \frac{1}{T_0}A(\omega){\rm e}^\omega\Big[ \beta \Bigl\{ I_0(\lambda)
          + \beta c_s\frac{\eta_{max}}{\lambda}I_1(\lambda) \Bigr\}\sinh \eta   \nonumber \\
        &&\hspace{1.8cm}    -(\beta c_s)^2  \frac{\eta}{\lambda}I_1(\lambda)\cosh \eta \Big].   \label{eq.opd5}        
 \end{eqnarray}
\section{Data Analysis}
\label{DA}

\begin{figure}[!htb]
   \begin{center}
      \includegraphics[width=7.8cm,clip]{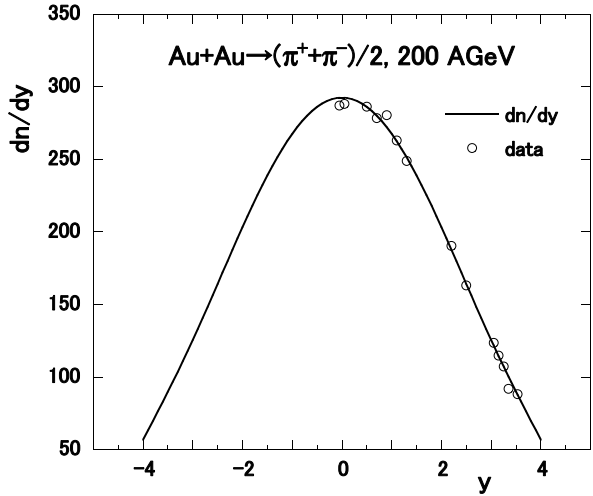}
   \end{center}
    \caption{\label{fig.pion}Analysis of single particle rapidity distribution of 
    charged pions $(\pi^+ \pi^-)/2$ by Eq.(\ref{eq.opd3}).}    
 \end{figure}

The single particle rapidity distribution of charged $\pi$ mesons in Au-Au collisions at 200 AGeV at RHIC~\cite{bear05} is analyzed 
by Eq.(\ref{eq.opd3}).  In the analysis, the initial temperature $T_0$ is fixed at 0.95 GeV, the critical temperature $T_c$ at 0.18 GeV
 and the freeze-out temperature $T_f$ at 0.12 GeV. The nuclear radius $R_A$ is parametrized as, 
 $R_A=1.2\times A^{1/3}$, where $A$ denotes the mass number of nucleus, and $A=197$ for Au. 
 Then, two parameters, $Q_0$ and $c_s$, remains in our formulation. Estimated parameters are shown in Table 1, 
 and the result on the single particle rapidity distribution of charged $\pi$ mesons is shown in Fig.~\ref{fig.pion}.

 \begin{table}[!htb]
  \begin{center}
     \caption{ Estimated values of parameters $Q_0$ and ${c_s}^2$  from Au+ Au $\rightarrow (\pi^+ + \pi^-)/2 + X$ }
   \begin{tabular}{ccc} \hline
       $Q_0$  &   ${c_s}^2$ & $\chi_{min}^2/{\text n.d.f}$ \\ \hline
       $(4.78 \pm 1.15)\times 10^{-2}$  &  $0.117\pm 0.005$ & 41.4/12 \\ \hline 
   \end{tabular}
  \end{center}
 \end{table}     

\begin{figure}[!htb]
  \begin{center}
      \includegraphics[width=7.8cm,clip]{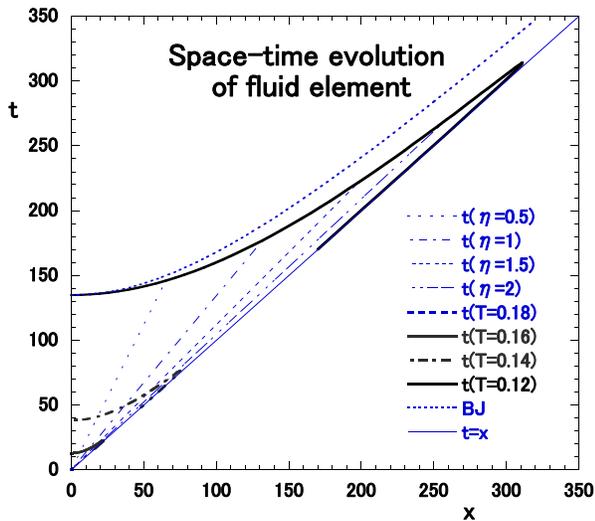}
  \end{center}
    \caption{\label{fig.xt-diaga} (Color online) Space-time evolution of fluid element at fixed temperature}
 \end{figure}

The space-time evolution of fluid element at $\omega$=constant in the $x$-$t$ plane is shown in Fig.\ref{fig.xt-diaga}. 
The profile at $T=T_c=0.180$ GeV is overlapped to the origin $(x,t)=(0,0)$. 
The dotted curve denoted by BJ in the figure means the Bjorken's scaling solution, $\tau=\sqrt{t^2-x^2}=$constant~\cite{bjor83}.

\begin{figure}[!htb]
   \begin{center}
      \includegraphics[width=7.8cm,clip]{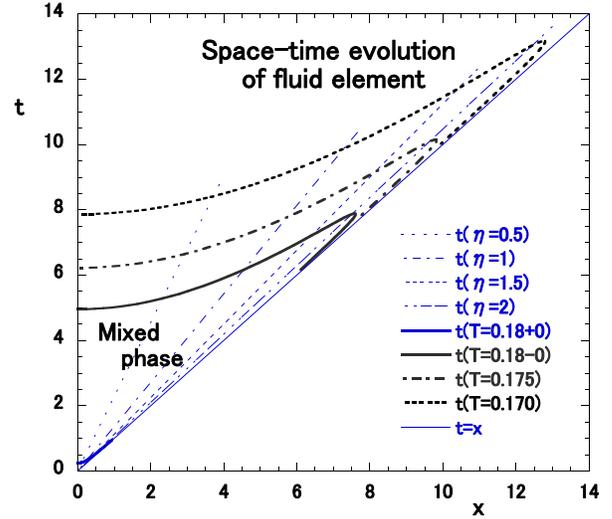}
  \end{center}
    \caption{\label{fig.xt-diagb} (Color online) Space-time evolution of fluid element at fixed temperature 
    from $T=0.180$ GeV to $T=0.170$ GeV.}
 \end{figure}

In order to investigate the space-time evolution of fluid element around $\omega=\omega_c$ or $T=T_c$,  
the detail of it is shown in Fig.\ref{fig.xt-diagb}.
The profile at $\omega=\omega_c-0$ or $T=T_c+0$ is in the QGP state. It is calculated by Eq.(\ref{eq.qgp8}). 
The profiles at  $\omega>\omega_c$ or $T<T_c$ are in the hadronic states.  
Those are calculated by Eqs.(\ref{eq.hdf2}) and (\ref{eq.hdf3}). 
 As can be seen from Fig.\ref{fig.xt-diagb}, the region between the profile at $T=T_c + 0$ and that at $T=T_c - 0$  
 corresponds to the mixed phase where the QGP state and the hadronic state are coexistent. 

%\newpage
%

\begin{figure}[!htb]
   \begin{center}
      \includegraphics[width=7.8cm,clip]{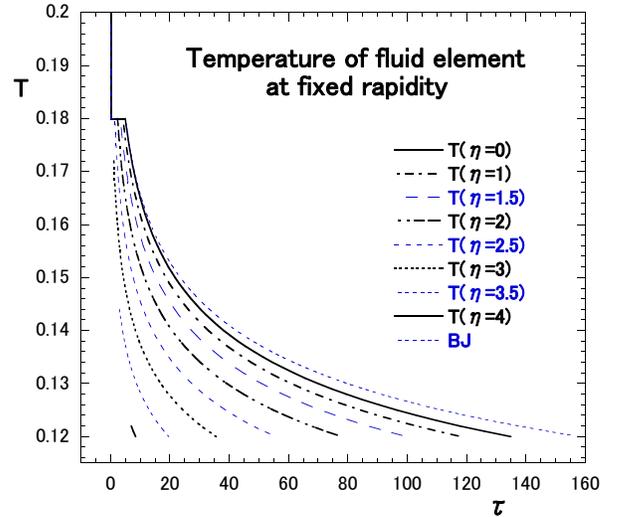}
  \end{center}
    \caption{ \label{fig.tau-temp} (Color online) Temperature profiles of fluid elements at fixed rapidity.}
 \end{figure}

The temperature dependence of fluid element at $\eta=$ constant is shown as a function of proper time, 
$\tau=\sqrt{t^2-x^2}$, in Fig.\ref{fig.tau-temp}. The temperature profiles of fluid elements for $T>0.200$ GeV are abbreviated.

The fluid element with larger rapidity in the absolute value is cooled faster, as can be seen from the figure.
The fluid starts expansion from $T=T_0$ at $\eta=0$ and at $\tau=0$.  
For example, the fluid element at $\eta=0$ is cooled down from $T=T_0$ at $\tau=0$ to $T=T_c$ at $\tau=0.238$ fm/c. 
It remains at $T=T_c$ from $\tau=0.238$ fm/c to $\tau=4.957$ fm/c. 
Therefore, the mixed phase in the neighborhood of $\eta=0$ is to continue about 4.7 fm/c. 
The fluid element at $\eta=0$ is cooled down to $T=T_f$ at $\tau=134.9$ fm/c.

 The dotted curve denoted by BJ corresponds to the Bjorken's scaling solution, $T=T_c(\tau_c/\tau)^{{c_s}^2}$ with ${c_s}^2=0.117$, 
 which is normalized to $T_c$ at $(\tau, T)$=$(\tau_c, T_c)$=$(0.238, 0.18)$ .

\section{Concluding Remarks}

We have formulated one dimensional hydrodynamical model including the phase transition from the QGP state to the hadronic state.
 At first, following to the Landau's hydrodynamical model, the equation of the telegraphy for the potential $\chi(\eta, \omega)$ is 
 solved under the simplified initial condition in the $\eta-\omega$ space with the velocity of sound, $c_0=1/\sqrt{3}$.  
 Then the solution of it in the hadronic state with the constant velocity of sound, $c_s$ ($c_s \neq c_0$), 
 is found so as to coincide with the solution in the QGP state if $c_s=c_0$. 
 
The space-time evolution of fluid element from $\omega=0$ (or $T=T_0$) to $\omega=\omega_f$ (or $T=T_f$) is calculated by our model. 
The non-zero finite region emerges between the profile at $\omega=\omega_c -0$ (or $T=T_c + 0$) and that at $\omega=\omega_c +0$ (or $T=T_c - 0$).  
This is due to the discontinuity of the potential $\chi(\eta, \omega)$ at $\omega=\omega_c$ (or $T=T_c$) 
caused by the change of the velocity of sound from $c_0$ in the QGP state to $c_s$ ($c_s < c_0$) in the hadronic state at $\omega=\omega_c$.
In our calculation, the fluid element at $\eta=0$ freezes out at $\tau=134.9$ fm/c.
 
 In the computer simulations of the three dimensional hydrodynamical models, 
 the evolution of fluid element is calculated as a function of the proper time $\tau$.  
 For example, the initial condition is taken as $\tau=0.6$ fm/c and $T_0=0.36$ GeV at 200 AGeV~\cite{kolb04}. 
 
 In our one dimensional hydrodynamical model, the initial condition is taken as $\tau=0$ fm/c and $T_0=0.95$ GeV.  
 The fluid element at $\eta=0$ is cooled down to $T=0.36$ GeV at $\tau=0.036$ fm/c. 
 Therefore, the temperature of fluid in our calculation at $\tau=0.6$ fm/c is not higher than 
 that of the three dimensional hydrodynamical model.
    
In the computer simulations at 200 AGeV, the fluid element expands a few hundred fm along the direction of colliding nuclei 
 before freeze-out though the value of $\tau$ at freeze-out is about 15 fm/c or so~\cite{mori07}. 
 
The proper time of fluid element at $\eta=0$ at freeze-out in our calculation is 
one order larger than the proper time of fluid element at freeze-out in the three dimensional computer simulation. 
In the one dimensional hydrodynamical model, the expansion of fluid into the transverse dimension is neglected 
contrary to the three dimensional calculations. 
Therefore, the temperature of the fluid element especially in the neighborhood of $\eta=0$ would decrease more slowly than 
that calculated by the three dimensional computer simulation, where the transverse expansion of the fluid is taken into account. 
However, the scale of expansion of fluid element in the space variable in our calculation shown in Fig.\ref{fig.xt-diaga} 
is comparable to that along the direction of colliding nuclei in the computer simulation.

In the recent lattice QCD calculations, it is shown that the crossover phase transition occurs 
at chemical potential $\mu=0$~\cite{aoki06, forc07}.
Let the crossover transition take place in the region,  ($\omega_c-\Delta\omega/2$, $\omega_c+\Delta\omega/2$). 
In our formulation, it would be expressed by the smooth change of the velocity of sound from $c_0$ to $c_s$.
The sound of velocity $c_v(\omega)$ in  ($\omega_c-\Delta\omega/2$, $\omega_c+\Delta\omega/2$)
is assumed to be a continuous function with positive value, 
and to satisfy $c_v(\omega_c-\Delta\omega/2)=c_0$ and $c_v(\omega_c+\Delta\omega/2)=c_s$. 

The potential $\chi(\omega)$ in ($0$, $\omega_c-\Delta\omega/2$) is the same with Eq.(\ref{eq.qgp8}).
The potential $\chi(\omega)$ after the crossover transition, in ($\omega_c+\Delta\omega/2$, $\omega_f$), 
would be given by the following equation,
 \begin{eqnarray}
     \chi(\eta,\omega) = A(\omega)I_0(\lambda),  \label{eq.cros1} 
 \end{eqnarray}
where,
 \begin{widetext}
  \begin{eqnarray}
             A(\omega) &=& \frac{Q_0c_0}{2}{\rm exp} \bigl[\beta(\omega-\omega_c-\Delta\omega/2)+\beta_0(\omega_c-\Delta\omega/2)
           +\int_{\omega_c-\Delta\omega/2}^{\omega_c+\Delta\omega/2}\beta_v(x) dx \bigr],
            \quad \beta_v(x) = \frac{1-{c_v(x)}^2}{2{c_v(x)}^2},  \label{eq.cros2}   \\
%  \end{eqnarray}
%
%
%  \begin{eqnarray}
     \lambda &=& \beta c_s \sqrt{\eta_{max}^2-\eta^2}, \quad  
          \eta_{max} = (\omega-\omega_c-\Delta\omega/2)/c_s +(\omega_c-\Delta\omega/2)/c_0
                      + \int_{\omega_c-\Delta\omega/2}^{\omega_c+\Delta\omega/2}\frac{1}{c_v(x)} dx.   \label{eq.cros3}      
  \end{eqnarray}
 \end{widetext}
\begin{acknowledgments}
The author would like to thank M. Biyajima, K. Morita and S. Muroya for valuable discussions.
\end{acknowledgments}

\bibliography{basename of .bib file}

\end{document}